\documentclass{aa}

\usepackage{natbib}
\usepackage{psfrag, color, epsfig}

\bibpunct{(}{)}{;}{a}{}{,}

\newcommand{\Vec}[1]{ {\boldmath\hbox{$#1$}\unboldmath} }

\newcommand{\average}[1]{\left\langle #1 \right\rangle}

\newcommand{\vp}{\varphi}
\newcommand{\vt}{\vartheta}
\newcommand{\ve}{\varepsilon}

\newcommand{\Cov}{\mbox{Cov}}

\newcommand{\dd}{{\mathrm d}}

\newcommand{\Map}{{M_{\rm ap}}}
\newcommand{\Mapsq}{M_{\rm ap}^2}
\newcommand{\qE}{q_{\rm E}}
\newcommand{\del}{\partial}
\newcommand{\DELTA}[3]{\Delta_{\vt_#1}^{#2 #3}}
\newcommand{\XI}[3]{\xi_#1(#2 #3)}
\newcommand{\tot}{{\rm tot}}

\newcommand{\Omegam}{\Omega_{\rm m}}

\newcommand{\ii}{{\rm i}}
\newcommand{\ee}{{\rm e}}
\newcommand{\gammat}{\gamma_{\rm t}}
\newcommand{\J}{{\rm J}}

\begin{document}

\title{Analysis of two-point statistics of cosmic shear: \\
II. Optimizing the survey geometry}

\author{Martin Kilbinger\inst{1} \and Peter Schneider\inst{1,2}
}

\institute{Institut f. Astrophysik
u. Extr. Forschung, Universit\"at Bonn, Auf dem H\"ugel 71,
D-53121 Bonn, Germany \and Max-Planck-Institut
f. Astrophysik, Postfach 1317, D-85741 Garching, Germany}

\offprints{Martin Kilbinger, \email{kilbinge@astro.uni-bonn.de}}

\date{Received / Accepted}

\abstract{We present simulations of a cosmic shear survey and show how
the survey geometry influences the accuracy of determination of
cosmological parameters.  We numerically calculate the full covariance
matrices $\Cov$ of two-point statistics of cosmic shear, based on the
expressions derived in the first paper of this series. The individual
terms are compared for two survey geometries with large and small cosmic variance.
We use analyses based on maximum likelihood of $\Cov$ and the Fisher
information matrix in order to derive expected constraints on
cosmological parameters. As an illustrative example, we simulate various
survey geometries consisting of 300 individual fields of $13^\prime
\times 13^\prime$ size, placed (semi-)randomly into patches
which are assumed to be widely separated on the sky and therefore uncorrelated.
Using the aperture mass statistics $\average{\Mapsq}$, the optimum
survey consists of 10 patches with 30 images in each patch. If $\Omegam,
\sigma_8$ and $\Gamma$ are supposed to be extracted from the data, the
minimum variance bounds on these three parameters are 0.17, 0.25 and
0.04 respectively. These variances raise slightly when the initial power
spectrum index $n_{\rm s}$ is also to be determined from the data. The
cosmological constant is only poorly constrained.  }

\titlerunning{Analysis of two-point statistics of cosmic shear II}
\maketitle

\section{Introduction}

Weak gravitational lensing by the large-scale matter distribution in the
Universe, called cosmic shear, has become a valuable tool for cosmology since
its first detection in 2000 \citep{2000MNRAS.318..625B, kaiser00, 2000A&A...358...30V, 2000Natur.405..143W}. Constraints on cosmological
parameters, in particular the (dark+luminous) matter density parameter $\Omegam$
and the power spectrum normalization $\sigma_8$, have been obtained from cosmic
shear with survey areas of up to several dozen square degrees
\citep[e.g.][]{J03, G02, vWMR01, vWM02, 2001A&A...368..766M, R02}.

Cosmic shear probes the statistical properties of the total matter
distribution projected along the line-of-sight between the observer and
distant galaxies which are typically at redshifts between 0.5 and 2. It
is independent of any possible bias between dark and luminous matter ---
e.g.\ no assumptions about how galaxies trace the dark matter have to be
made. 

Cosmic shear is sensitive to a large number of cosmological parameters, most notably
on $\Omegam, \sigma_8$ and the shape parameter $\Gamma$, but also 
to the source galaxy redshift distribution. The dependancy on these parameters is
partially degenerate.
Some of these near-degeneracies can be broken when weak lensing is combined with other
cosmological measurements like CMB anisotropy experiments, the
statistics of the Lyman-$\alpha$ forest or galaxy redshift surveys. The
parameter dependencies are very different for the individual methods,
for example the $\Omegam$-$\sigma_8$-degeneracy is nearly orthogonal
bet\-ween cosmic shear and CMB \citep{vWM02}. Even the
most precise measurement of cosmological parameters up to now, which
comes from the first data release of WMAP \citep{Bennett03, Spergel03},
 can be improved substantially when weak lensing data is
added \citep{Contaldi03, HuTeg99}.

In the first paper of this series \citep[ hereafter Paper I]{SvWKM02},
we reviewed the properties and relations of various two-point
statistics of cosmic shear. We defined unbiased estimators and
calculated their covariances. In Sect.\ 5 of Paper I, an explicit
expression for the covariance of the two-point correlation function was
given. Using that, we calculated expected constraints on various
cosmological parameters for a cosmic shear survey.

However, this ansatz is an approximation which is only valid for a
large and connected survey area.  Any real cosmic shear survey will
most likely consist of single unconnected
fields-of-view and have a complicated geometry. In this paper, we
present a method which allows one to calculate the covariance of the
two-point functions of cosmic shear for an arbitrary survey geometry.

The measurement of cosmic shear with a sufficient high precision
to constrain cosmological parameters requires many
independent lines-of-sight, lowering the sampling variance
(``cosmic variance''). On the other hand, it is important to measure
the shear on a large range of angular scales.  Even with modern
wide-field imaging cameras, separations of more than a few degrees cannot be
accessed by individual fields-of-view --- one has to observe some
fields near to each other and measure galaxy shape
correlations across individual fields.  In a recent work \citep{J03},
cosmic shear has been measured from 1 to 100 arc minutes, and
constraints on $\Omegam$ and $\sigma_8$ have been obtained.

\citet{1998ApJ...498...26K} remarked that for a survey of 9 square
degrees, consisting of a single $3^\circ \times 3^\circ$-field, the noise
due to the intrinsic ellipticity dispersion can be neglected on large
scales, because of the huge number of galaxy pairs. He also noted that
cosmic variance is dominant, and that ``sparse sampling'', meaning the
distribution of smaller fields on a larger region of the sky, reduces
the cosmic variance dramatically.

A cosmic shear survey has to cover a large area containing hundreds of
thousands of galaxies, whose shapes can be determined. Because
telescope time is limited, one has to carefully choose the locations
of the pointings, in other words, the geometry of the survey.
In this work, cosmic shear surveys with different geometrical configurations
are simulated. They are compared with respect to their
ability to constrain cosmological parameters, using two-point statistics.

The individual fields-of-view of the simulated surveys are placed in patches on
the sky in order to measure the shear correlation on large angular scales.
Several patches, distributed randomly on the sky in order to reduce cosmic
variance, build up the survey.

The comparison of the geometries is done with a likelihood analysis
using the covariances of two-point statistics of cosmic shear. These were
derived in Paper I, Sects.\ 4 and 6.  In this paper, the covariance
matrices are calculated via a Monte-Carlo-like method using the
simulated galaxy positions of the surveys.

The different survey geometries considered here are presented in Sect.\
\ref{sec:survey-geom}.  In Sect.\ \ref{sec:rev}, we review the
two-point statistics of cosmic shear relevant for this work, as well
as their estimators and covariances, as derived in Paper I.  The
method for the numerical calculation of the covariances is given in
Sect.\ \ref{sec:num}. Results for some patch geometries are presented
in Sect.\ \ref{sec:comp}. A likelihood analysis is performed in Sect.\
\ref{sec:likelihood}, where the expected constraints on pairs of
cosmological parameters is considered and compared for various cosmic
shear survey geometries. Finally, in Sec.\ \ref{sec:fisher}, the Fisher
information matrix is used to compare constraints on three and four parameters
simultanously.

This work is intended to be a preparation for a cosmic shear survey with the
wide-field camera VIMOS on the 8.2m ESO-VLT telescope UT3 (Melipal). The
numerical codes used for this paper are publically available\footnote{
\texttt{http://www.astro.uni-bonn.de/$\,\tilde{}\,$kilbinge/cosmicshear}}.

\section{Survey geometries}
\label{sec:survey-geom}

The different survey geometries considered in this work consist of
circular patches, in which individual images are distributed randomly,
but non-overlapping.  We define ``image'' as one single field-of-view.  It is
assumed that the shear correlation functions can be measured across
image boundaries, thus the cosmic shear can be determined in principle
on scales up to the patch diameter. Because there are always bright
stars or foreground galaxies which have to be avoided, we cannot
specify in much more detail the image positions, thus a random
distribution of the images in a patch is assumed.

The patches are assumed to be separated by at least several
degrees. On these scales, the correlation functions are virually zero,
so different patches can be considered as uncorrelated; they probe
statistically independent parts of the large-scale structure.

In our simulations, the individual images are $13^\prime \times
13^\prime$-fields, corresponding roughly to the field-of-view of VIMOS.
A survey consists of $P$ patches of
radius $R$, each patch containing $N$ images. The total number of
images, $n = P \cdot N$ is kept fixed for all geometries. The larger
the number of patches $P$, the smaller is the cosmic variance. On the
other hand, the larger the number of images $N$ per patch, the larger
is the number of galaxy pairs for which the correlation is measured,
thus the lower is the shot noise.  One of the goals of this work is to find
a configuration which is optimal in the sense that the two-point
correlation function can be measured most accurately; we characterize
this `accuracy' by considering constraints on pairs of cosmological
parameters from the measurement of the correlation function.

\begin{figure}[tb]
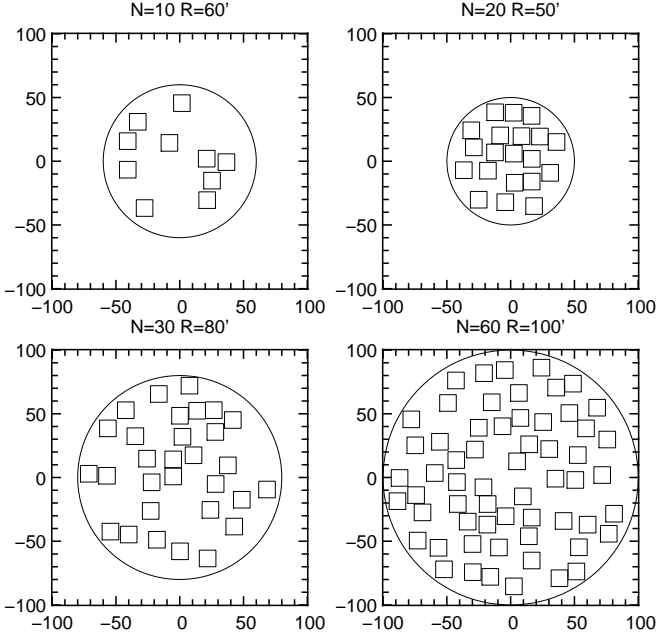

\resizebox{\hsize}{!}{
\includegraphics{np10-060_000-geom.ps}
\includegraphics{np20-050_002-geom.ps}}
\resizebox{\hsize}{!}{
\includegraphics{np30-080_000-geom.ps}
\includegraphics{np60-100_002-geom.ps}}
\caption{Realisations of patches; the squares represent the individual images. $N$
is the number of images per patch, $R$ the patch radius. Each image
has a size of 13 arc minutes.
}
\label{fig:patches}
\end{figure}

We use a total image number of $n=300$, corresponding to 14 square
degrees for the whole survey. For $N$, being a factor of $n$,
the values 10, 20, 30, 50, and 60 are considered, corresponding to geometries
with $P=$ 30, 15, 10, 6 and 5 patches, respectively. An illustration of some
patches is given in Fig.\ \ref{fig:patches}.

The patch geometries are compared to a configuration which consists of
300 single uncorrelated images, where `uncorrelated' again means
separated by at least several degrees. This
configuration has the smallest possible cosmic variance, but the shear
correlation can be only measured up to
$\sqrt{2} \cdot 13$ arc minutes.

\section{Two-point statistics of cosmic shear, their estimators and covariances}
\label{sec:rev}

We follow here the notation of \citet{BS01}.
See Paper I for a more detailed presentation of the formulae reviewed in this Section.
Two points $\vec \vt$
and $\vec \vt + \vec \theta$ define in a natural way a direction given by the connecting
vector $\vec \theta$, with respect to which the tangential and cross-component of the
shear,
\begin{equation}
\gamma_{\rm t} \equiv - \, \Re \left( \gamma \ee^{-2 \ii \vp} \right) \quad\quad \mbox{and} \quad \quad
\gamma_\times \equiv - \, \Im \left( \gamma \ee^{-2 \ii \vp} \right)
\label{gamma-tx-def}
\end{equation}
are defined, where $\vp$ is the polar angle of the connecting vector.
The two independent shear correlation functions $\xi_\pm(\vt)$ are
related to the power spectrum of the projected
matter density $P_\kappa$,
\begin{eqnarray}
\xi_{\pm}(\theta) & \equiv & \average{\gammat(\Vec \vt) \gammat(\Vec \vt +
\Vec \theta)} \pm \average{\gamma_\times(\Vec \vt) \gamma_\times(\Vec
\vt + \Vec \theta)} \label{xi-gg-def}\\
& = & \frac{1}{2 \pi} \int\limits_0^\infty \dd \ell
\, \ell \, P_\kappa(\ell) \J_{0,4}(\ell \theta),
\label{xi-pm-def}
\end{eqnarray}
where the first-kind Bessel function $\J_0$ ($\J_4$) corresponds to the `$+$'
(`$-$') correlation function, see e.g.\ \citet{1992ApJ...388..272K}. Another second-order
statistics of cosmic shear is the dispersion of the weighted tangential
shear in a circular aperture $B$ of radius $\theta$ \citep{S96},
\begin{equation}
\Map(\theta) = \int\limits_{B(\theta)} \dd^2 \vt \, Q(\vt) \gamma_{\rm
t}(\vt).
\label{Map-def}
\end{equation}
With the weight function $Q(\vt) = {2}/{\vt^2} \int_0^\vt \dd \rho \,
\rho \, U(\rho) - U(\vt)$, where $U$ is a compensated filter function,
i.e.\ $\int_0^\theta \dd \vt \, \vt \, U(\vt) = 0$ \citep{1998MNRAS.296..873S},
 its dispersion is related to the power spectrum by
\begin{equation}
\average{ \Mapsq(\theta)} = \frac{1}{2\pi} \int\limits_0^\infty \dd \ell \, \ell \,
P_\kappa(\ell) \left( \frac{24 \, \J_4(\ell \theta)}{( \ell \theta )^2}
\right)^2 .
\label{Map-disp}
\end{equation}
This aperture mass measures the gradient component or E-mode of the
shear.  Analogously to (\ref{Map-def}), one can also define the
weighted cross-component of the shear in an aperture,
$\average{M_\perp}$, which is a measure of the curl component
or B-mode only.

The two aperture mass statistics can be expressed in terms of the two-point
correlation functions,
\begin{eqnarray}
\lefteqn{\average{M_{{\rm ap}, \perp}^2(\theta)} =} \nonumber \\
 & & \frac{1}{2 \theta^2}
\int\limits_0^{2 \theta} \dd \theta^\prime \, \theta^\prime \, \left[
\xi_+(\theta^\prime) T_+\left( \frac{\theta^\prime}{\theta} \right) \pm 
\xi_-(\theta^\prime) T_-\left( \frac{\theta^\prime}{\theta} \right) \right],
\label{Map-xi-EB}
\end{eqnarray}
where $T_+$ and $T_-$ are given explicitly in
\citet{2002A&A...389..729S}.

The shear due to the tidal gravitational field of the large-scale
structure is a pure gradient field\footnote{at least in first
approximation}
\citep{1995ApJ...439L...1K}, therefore, no B-modes should be present. However,
there are other effects which can produce B-modes; these are mainly
systematic measurement errors and intrinsic galaxy orientation
correlations. The contribution from the latter can be reduced if the
survey is deep, or if photometric redshift information of the source
galaxies is taken into account \citep{KS02, HH03}.  Minor
contributions to B-modes are source clustering
\citep{2002A&A...389..729S} and higher-order lensing effects
\citep{JSW00}.

The separate measurement of E- and B-modes allows one to quantify
contaminations to the gravitational shear signal. In recent cosmic
shear surveys, a non-zero B-mode signal has been measured
\citep{vWM02, hoe02, J03},
using the statistics (\ref{Map-xi-EB}). However, a recent re-analysis
of the VIRMOS-DESCART data shows no significant B-mode signal any
more; the previously found B-modes were obviously due to insufficient
PSF corrections\footnote{van Waerbeke, priv.\ comm.}.

\subsection{Shear estimators}
\label{sec:shear-estim}

The shear estimators used here are similar to those introduced in
Paper I. For simplicity, all weight factors which account for
differences in the precision of ellipticity measurement of individual
galaxies are set to unity.  Further, we assume that the correlation
function is to be estimated in logarithmic bins; therefore, the
following function is defined,
\begin{equation}
\Delta_\vt(\theta) = \left\{ \begin{array}{ll} 1 \;\;\;\;  & \mbox{for}
\;\;
\left | \ln \vt - \ln \theta \right| < \frac{\Delta \ln \theta}{2} \\
0 \;\;\;\; & \mbox{otherwise} \end{array}
\right.,
\end{equation}
which selects bins with logarithmic bin width $\Delta \ln \theta$
around $\vt$. Then, if the $i$-th galaxy is located at angular
position $\vec
\theta_i$ with the observed ellipticity $\ve_i$,
\begin{equation}
\hat \xi_{\pm}(\vt) = \frac{1}{N_{\rm p}(\vt)} \sum_{ij} ( \ve_{i {\rm t}} \ve_{j {\rm t}} \pm
\ve_{i\times} \ve_{j\times} ) \Delta_\vt(|\Vec \theta_i - \Vec
\theta_j|),
\label{xi-hat}
\end{equation}
are unbiased estimators of the correlation functions
(\ref{xi-gg-def}). Here, $N_{\rm p}(\vt)$ is the number of galaxy
pairs in the bin corresponding to $\vt$.  The double sum is performed
over all galaxy pairs. 

Unbiased estimators for the aperture mass dispersions $\average{\Mapsq}$
and $\average{M_\perp^2}$ are
\begin{eqnarray}
\lefteqn{ {\cal M}_\pm(\theta) = } \nonumber \\
& & \frac{1}{2 \theta^2} \sum_{i=1}^{I} \Delta \vt_i
\vt_i \left[ \hat \xi_+(\vt_i) T_+\left( \frac{\vt_i}{\theta} \right) \pm 
\hat \xi_-(\vt_i) T_-\left( \frac{\vt_i}{\theta} \right) \right],
\label{Map-est}
\end{eqnarray}
where an index has been attached to the bin width $\Delta \vartheta_i$ to account
for variable (e.g. logarithmic) bin widths. The limit $I$ of the sum must
be chosen such that $\theta$ is half the upper limit of the $I$-th bin.

\subsection{Covariances}
\label{sec:cov}

In Paper I we calculated the covariance matrices of the estimators
defined in the last section. These consist of several terms which we
call shot noise or diagonal term ($D$), mixed term ($M$) and pure cosmic variance
term ($V$). We perform the following decomposition:
\begin{eqnarray}
\Cov(\hat \xi_+, \vt_1 ; \hat \xi_+, \vt_2) & = & D \, + \, M_{++} \, + \, V_{++}
\nonumber \\
\Cov(\hat \xi_-, \vt_1 ; \hat \xi_-, \vt_2) & = & D \, + \, M_{--} \, + \, V_{--}
\label{Cov} \\
\Cov(\hat \xi_+, \vt_1 ; \hat \xi_-, \vt_2) & = & M_{+-} \, + \, V_{+-}, \nonumber
\end{eqnarray}
where the individual terms are
\begin{eqnarray}
D & \equiv & \frac{\sigma_\ve^4}{F} \bar
\delta(\vt_1 - \vt_2) N_{\rm p}(\vt_1) \nonumber \\
M_{++} & \equiv & \frac {2 \sigma_\ve^2} F \,
 \sum_{ijk} \DELTA 1 i j \DELTA 2 i k \XI + j k \nonumber \\
V_{++} & \equiv & \frac 1 F \sum_{ijkl} \DELTA 1 i j \DELTA 2 k l \Big( \XI + i l \XI +
j k \nonumber \\
& & + \cos 4 (\vp_{il} - \vp_{jk}) \XI - i l \XI - j k \Big) \nonumber\\
M_{--} & \equiv & \frac{2 \sigma_\ve^2} F \,  \sum_{ijk} \DELTA 1 i j \DELTA 2
i k \cos 4 ( \vp_{ik} - \vp_{ij} ) \XI + j k \label{Cov-terms} \\
V_{--} & \equiv & \frac 1 F \sum_{ijkl} \DELTA 1 i j \DELTA 2 k l \Big( \cos 4
(\vp_{ij} - \vp_{il} - \vp_{jk} + \vp_{kl} ) \nonumber \\
& & \hspace*{-0.6cm} {} \times \XI - i l \XI - j k +
\cos 4 ( \vp_{ij} - \vp_{kl} ) \XI + i l \XI + j k \Big) \nonumber \\
M_{+-} & \equiv & \frac{2\sigma_\ve^2} F \, \sum_{ijk} \DELTA 1 i j \DELTA 2
i k \cos 4 ( \vp_{ik} - \vp_{jk} ) \XI - j k \nonumber \\
V_{+-} & \equiv & \frac 2 F \sum_{ijkl} \DELTA 1 i j \DELTA 2 k l
\cos 4 ( \vp_{il} - \vp_{kl} ) \XI - i l \XI + j k, \nonumber
\end{eqnarray}
with $F \equiv {N_{\rm p}(\vt_1)N_{\rm p}(\vt_2)}$,
$\Delta_\vt^{ij} \equiv \Delta_\vt(\left| \vec \theta_i - \vec \theta_j
\right|)$ and $\xi_\pm(ij) \equiv \xi_\pm(\left| \vec \theta_i - \vec \theta_j\right|)$
for brevity. $\bar \delta(\vt_1 - \vt_2)$ is unity if $\vt_1$ and
$\vt_2$ are in the same bin and zero otherwise. $\sigma_\ve$ is the
ellipticity dispersion of the galaxies in the absence of shear.

The covariances of the aperture mass dispersions are
\newlength{\xxx}
\setlength{\xxx}{6em}
\begin{eqnarray}
\Cov({\cal M}_\pm; \theta_1, \theta_2) & = & \frac{1}{4 \, \theta_1^2
\theta_2^2} \sum_{i=1}^{I_1} \sum_{j=1}^{I_2} \Delta \vt_i \Delta \vt_j
\vt_i \vt_j \nonumber \\
& & \hspace*{-\xxx} {}  \times  \Bigg[ T_+ \left( \frac{\vt_i}{\theta_1} \right)
T_+ \left( \frac{\vt_j}{\theta_2} \right) \Cov(\hat \xi_+, \vt_i; \hat
\xi_+, \vt_j) \nonumber \\
& & \hspace*{-\xxx} {} + T_- \left( \frac{\vt_i}{\theta_1} \right)
T_- \left( \frac{\vt_j}{\theta_2} \right) \Cov(\hat \xi_-, \vt_i; \hat
\xi_-, \vt_j) \nonumber \\
& & \hspace*{-\xxx} {} \pm T_+ \left( \frac{\vt_i}{\theta_1} \right)
T_- \left( \frac{\vt_j}{\theta_2} \right) \Cov(\hat \xi_+, \vt_i; \hat
\xi_-, \vt_j) \nonumber \\
& & \hspace*{-\xxx} {} \pm T_+ \left( \frac{\vt_j}{\theta_2} \right)
T_- \left( \frac{\vt_i}{\theta_1} \right) \Cov(\hat \xi_+, \vt_j; \hat
\xi_-, \vt_i) \Bigg].
\label{CovMap-def}
\end{eqnarray}
The upper limits $I_k, \; k=1,2,$ must be chosen such that $\theta_k$ is half the
upper limit of the $I_k$-th bin. For brevity, the notions $\Cov_{++} \equiv
\Cov(\hat \xi_+, .; \hat \xi_+, .),
\Cov_{--} \equiv \Cov(\hat \xi_-, .; \hat \xi_-, .)$ and $\Cov_{+-}
\equiv \Cov(\hat \xi_+, .; \hat \xi_-, .)$ are used from now on.

We note here that the expressions for the cosmic variance terms $V$
of the covariances are
only valid if the shear field is Gaussian. On scales below $\sim$ 10
arc minutes, the non-Gaussianity of the shear field gets important,
e.g.\ Fig.\ 4 of
\cite{vWM02}, see also \cite{1999ApJ...527....1S}. On scales below 1 arc
minute, the shot noise term $D$ dominates over $V$, see Fig.\ \ref{fig:diag}
of this paper and Fig.\ 3 of Paper I, thus with the Gaussian assumption we expect to
slightly underestimate the covariances in this transition regime between 1 and
10 arc minutes.

\section{Numerical calculation of the covariance matrices}
\label{sec:num}

Given a model for the shear correlation functions or, equivalently, the power
spectrum $P_\kappa$, the covariance matrices
(\ref{Cov})
only depend on the positions of the
observed galaxies, in other words, on the survey geometry.  For a
given data set with known positions of observed galaxies, it is
straightforward to calculate the covariances. An \`a
priori estimate of the covariances is made using simulated galaxy positions
for the summations in (\ref{Cov-terms}). Note that only the positions
of the galaxies have to be simulated, not their ellipticities. In
order not to introduce artificial Poisson noise, the galaxies are not distributed
randomly but subrandomly onto the fields, see Chapter 7 of 
\citet{nr}.

Throughout, 20 logarithmic bins in angular separation are used, the
smallest bin being centred around 10 arc seconds. 
The largest separation considered is either $\sqrt 2 \cdot 13^\prime$
for the uncorrelated images, or equal to the patch radius $R$ in the
case of a patch geometry. Thus, the bin widths differ for geometries with different patch radii.

\begin{figure}[tb] \resizebox{\hsize}{!}{\input{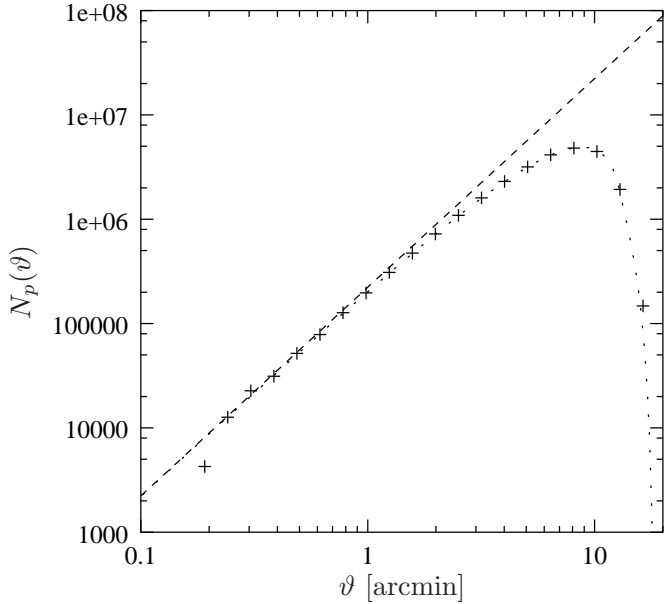}}
\caption{Number of galaxy pairs per angular bin $N_{\rm p}$ against the
bin centre $\vt$, for a single $13^\prime \times 13^\prime\,$-field. The
points are the pair numbers obtained from the subrandom distribution,
the dashed line is the approximation $N(\vt) = 2 \pi A n^2 \vt \Delta
\vt$ (Paper I), where $A$ is the survey area and $n$ the galaxy number
density, which was taken to be 30 per square arc minute. The dotted
curve is from \protect\citet{KS03}.} \label{fig:Npair} \end{figure}

The number of galaxy pairs per bin is shown in Fig.\ \ref{fig:Npair} for a
single $13^\prime
\times 13^\prime\,$-field. For intermediate angles, $N_{\rm
p}(\vt) \propto \vt^2$, as follows from eq.\ (26) of Paper I, with $\Delta \vt
\propto \vt$ for logarithmic bins. Deviations show up for large scales, which are
due to boundary effects and for very small scales, because of the
subrandom galaxy distribution. The curve agrees well with the
theoretical expectation \citep{KS03}.

Throughout, we set the intrinsic ellipticity dispersion to $\sigma_\ve = 0.3$.
The number of galaxies per square arc minute, for which a shape mesurement is
feasible, is set to 30. To obtain this number density, a limiting $R$-band magnitude of
about 25.5 is needed. 

\label{sec:pair-sel}

Because the number of operations for the calculation of the
covariances increases with the number of galaxies to the fourth power,
it is not feasible, except for a very small survey area, to sum over
all galaxy positions. Instead, a random subsample of galaxies is used.

The summations over galaxy positions can be written as sums over pairs
of galaxies, which have a fixed separation for each matrix element
$(\vt_1, \vt_2)$, as determined by the $\Delta$-functions. Thus it is
convenient to store galaxy pairs for each angular bin.  The mixed terms
$M$ (\ref{Cov-terms}) can then be split up into a sum over all those $\vt_1$- and $\vt_2$-pairs
which have a galaxy in common. This can be done efficiently if the
pairs for each bin are sorted by galaxy number. The cosmic variance terms $V$ are
simply double sums over all $\vt_1$- and $\vt_2$-pairs.

In order to decrease the computing time to a feasible value, we use random
subsamples of all galaxy pairs for the summations, which consists of $300\,000$
pairs per bin for the triple sums and $1\, 000$ pairs per bin for the quadruple
sums.

The calculation of the individual addends in the covariances is
straightforward. The correlation functions are obtained by linear
interpolation between grid points calculated beforehand, using
(\ref{xi-pm-def}). The cosine of the sum of angles is expanded into
a sum over products of cosines, and by using the relation
$\cos 4 \vp = 2 [ 2 x^2/(x^2+y^2)^2 - 1]^2 - 1$, where $\vp$ is
the polar angle of the vector $(x,y$), no single
time-consuming trigonometric function actually has to be evaluated.

\subsection{Cosmological model}
\label{sec:cosm}

For the power spectrum of the matter fluctuations, we assume an initial
power spectrum $P_{\rm i} \propto k^{n_{\rm s}}$, the transfer function for Cold
Dark Matter from \cite{bbks86} and the fitting formula for
the non-linear evolution of \citet{pd96}. The redshift
distribution of the source galaxies is \citep{Smail95}.
\begin{equation}
p(z) \dd z = \frac{\beta}{z_0 \Gamma\left(3/\beta\right)} \left(
\frac{z}{z_0} \right)^2
\ee^{- \left( z/z_0 \right)^\beta} \dd z,
\label{prob-def}
\end{equation}
where $\Gamma$ denotes the Eulerian gamma function.  Our reference
cosmology is a flat $\Lambda$CDM model with $\Omegam = 0.3$,
$\Omega_\Lambda = 0.7$, the shape parameter $\Gamma=0.21$, $n_{\rm s}
= 1$ and the normalisation $\sigma_8=1$. The parameters of the
redshift distribution are $z_0=1$ and $\beta=1.5$, which corresponds
to a mean source redshift of $\approx 1.5$.

\subsection{The covariance matrices}
\label{sec:cont}

\begin{figure*}
\begin{center}
\input{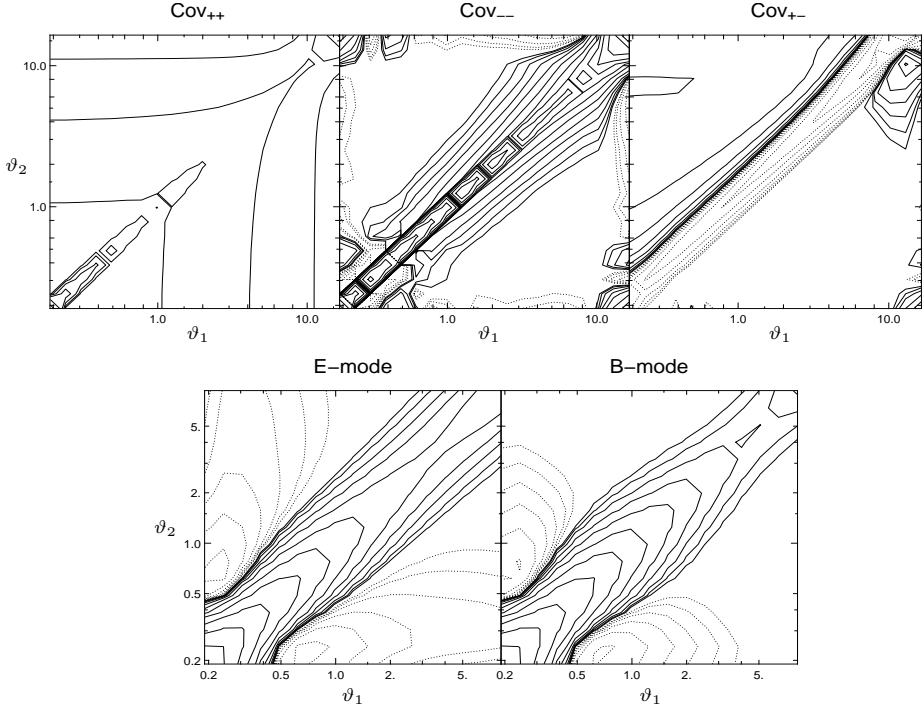}
\end{center}
\caption{Contour plots of the covariance matrices of the correlation
functions (\ref{Cov-terms}) and the aperture mass dispersions
(\ref{CovMap-def}) $\average{\Mapsq}$ (``E-mode'') and
$\average{M_\perp^2}$ (``B-mode''), for a single $13^\prime \times
13^\prime\, $-image.  Logarithmic contours are used, ranging from $2
\times 10^{-13}$ to $10^{-8}$, with 4 spacings per decade. Negative
values are plotted as dotted contours. The `boxiness' of some contour
lines is due to the coarse resolution; the matrices are calculated on
a logarithmic $20
\times 20 \, $-grid for $\vt_1$ and $\vt_2$.  The axes show $\vt_1$
and $\vt_2$ in arc minutes.  Note that the aperture mass covariances can only be
calculated up to half the largest scale where the covariances of the
correlation functions are known.}
\label{fig:13-cov-contours}
\end{figure*}

In Fig.\ \ref{fig:13-cov-contours}, contour plots of the three
covariance matrices are shown for a single $13^\prime \times
13^\prime\,$-field.  The diagonal of $\Cov_{++}$ is enhanced above the
non-diagonal elements only for small angular separations, which is
mainly due to the $(1/N_p)$-dependence of the shot noise term $D$.
The $\Cov_{--}$-matrix shows a more contrasted diagonal, with a
rapid falloff away from the diagonal. This stems from the fact that $\xi_-$
filters the convergence power spectrum more locally than $\xi_+$,
resulting in smaller intercorrelations of different angular
scales. The asymmetric matrix $\Cov_{+-}$ has also negative diagonal
elements, meaning anticorrelation. For $\vt_2 \approx 1.4 \, \vt_1$,
$\Cov_{+-}$ is zero.

The covariance matrices of the two aperture mass statistics are quite
similar to each other, indicating that the third and fourth terms in
(\ref{CovMap-def}) are small in comparison to the first two terms.
Moreover, they also resemble $\Cov_{--}$, which is due to the resembling
functional behaviour of $\xi_-$ and $\average{\Mapsq}$; the filter
functions for both statistics ($J_4(x)$ and $[J_4(x)/x]^2$
respectively) are strongly peaked and thus filter the power spectrum
very locally.

\subsection{Comparison of various geometries}
\label{sec:comp}

\begin{figure*}[ht!]
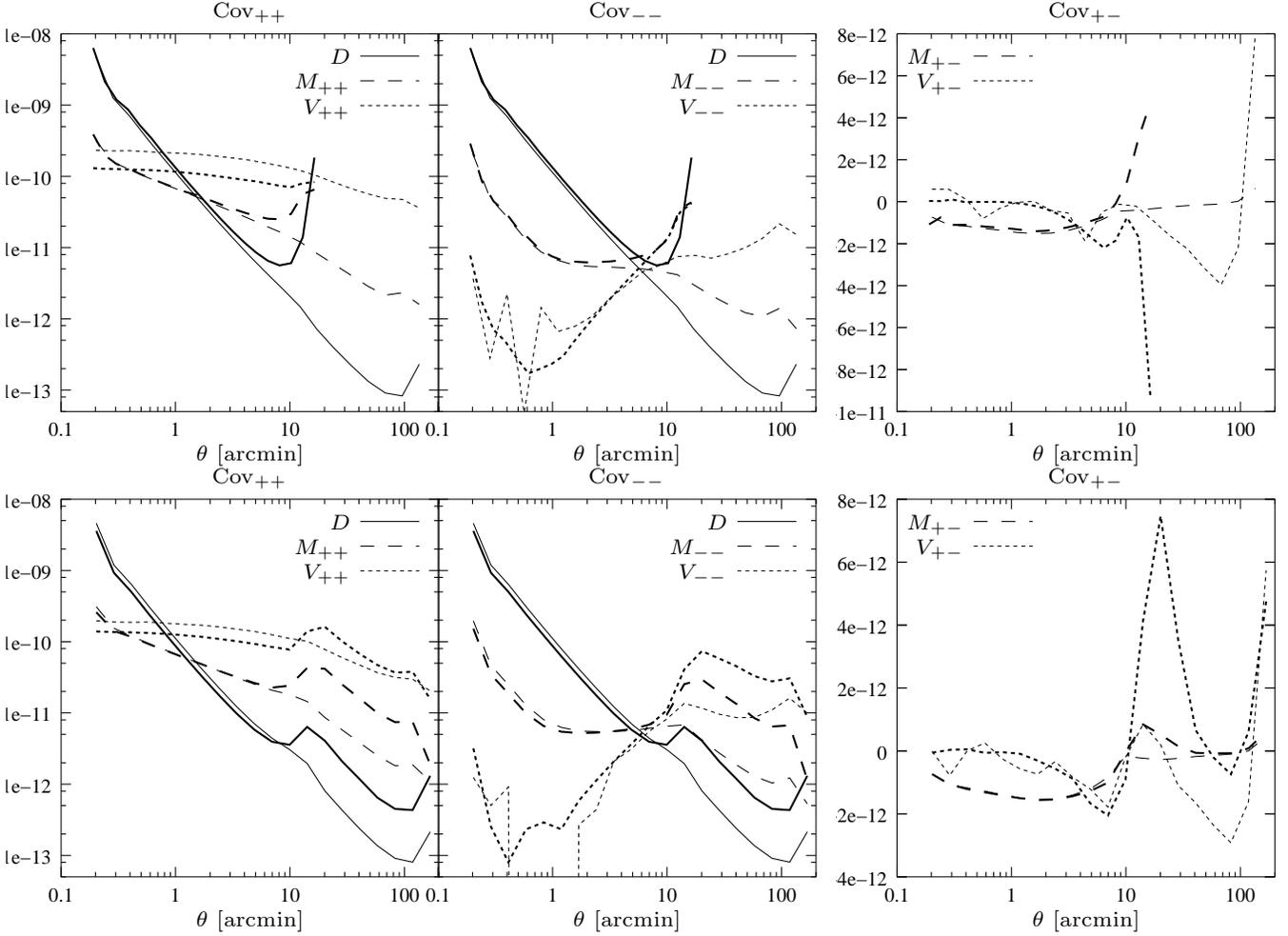

\resizebox{\hsize}{!}{
\input{diag++--.pstex_t}
\input{diag+-.pstex_t}}
\resizebox{\hsize}{!}{
\input{diagy++--.pstex_t}
\input{diagy+-.pstex_t}}
\caption{The diagonal elements of the covariance
matrices (\ref{Cov}), split up into the individual
terms $D$, $M$ and $V$ (\ref{Cov-terms}).
\emph{Upper row:} 300 single uncorrelated images, where
the largest scale is $\sqrt{2} \cdot 13^\prime$ (thick lines)
and a patch geometry with $N=60$ and $R=80^\prime$ (thin
lines). \emph{Lower row:} $N=10$, $R=100^\prime$ (thick lines) and $N=60$,
$R=100^\prime$ (thin lines). The left and middle panels
are logarithmic plots, the $y-$axis of the right panel is linear.
Note that the bin widths are different  for the different geometries
which affects $D$. Because logarithmic bins are used, the
differences in the bin widths are small, however.}
\label{fig:diag}
\end{figure*}

In Fig.\ \ref{fig:diag}, the individual terms (\ref{Cov-terms})
contributing to the diagonal ($\vt_1 = \vt_2$) of the covariance
matrices (\ref{Cov}), calculated for four survey geometries, are compared. One
sees the $(1/N_{\rm p})$-dependence of $D$; it decreases as the
number of pairs increases towards larger separations. Only for scales
comparable to the image or patch boundary, the number of pairs
decreases. This leads to an increase of $D$ (see also Fig.\
\ref{fig:Npair}). For $\,
\Cov_{++}$, the cosmic variance $V$ dominates on scales larger than about
one arc minute. $\Cov_{--}$ is dominated by $D$. The
$\Cov_{+-}$-elements are very small compared to those of the two
symmetric matrices.

In the upper row of Fig.\ \ref{fig:diag}, all terms for the two
extreme geometries regarding cosmic variance are shown: the 300
uncorrelated images, and the configurations with only five patches with a
radius of 80 arc minutes. The difference is about a factor of two in
the cosmic variance term $V_{++}$ The other terms are quite similar
for the two geometries, except on scales comparable to the image size.
 $V_{--}$ is much smaller; at this level, the differences
between the geometries are presumably mainly due to numerical noise,
as well as the negative value at about 0.6 arc minutes for the patch geometry
. Thus, $\xi_+$ is much more affected by cosmic variance than $\xi_-$.

Obviously, the covariances corresponding to the patch geometry extend
to larger angular scales than those for the uncorrelated images.
One important question which is adressed in
this paper is whether the additional information of the shear on
large scales can compensate for the larger cosmic variance on smaller
scales.

The lower row of panels of Fig.\ \ref{fig:diag} compares the
covariance terms of patch geometries with the same radius, but with a
low ($N=10$) and a high ($N=60$) image density in the patches. In the
first case, there is quite a sharp transition at a scale where the
image boundary is exceeded, all terms increase at about 10 arc
minutes. The case $N=60$ shows a less drastic change; because of the
higher image density, a large number of pairs at this separation on
different images is found, thus the transition is smoother.

\subsection{Comparison with the approximation of Paper I}
\label{sec:method}

We compare the single terms of the covariance matrix $\Cov_{++}$
(\ref{Cov}) obtained by the summation of simulated galaxy positions
presented in this paper with the integration method from Paper I,
Sect.\ 5, for a large connected field.  We found deviations at large
separations, where the assumption validating the approximation for the
integration method breaks down. Further, for small separations, the
diagonal elements of the mixed term $M_{++}$ term are enhanced in the
summation method. This is due to the discreteness of the galaxies: For
$\vt_1 = \vt_2$, there are $N_{\rm p}(\vt_1)$ summands where
$j=k$. This gives a contribution of $N_{\rm p}(\vt_1) \, \xi_+(0)$
which is not present for the off-diagonal elements.  A similar but
smaller contribution adds to $V_{++}$. These effects do not enter the
integration method, where a smooth galaxy distribution is assumed.

Further differences between the two methods were found for the
cosine-part of the $V_{++}$-term which is due to fact that $\xi_-$
decreases very slowly for large separations and thus the
$\vp$-integration in eq.\ (34) of Paper I obtains a considerable
contribution from separations which are larger than the field
boundary.

Altogether, the deviations are quite small. In particular, the
resulting likelihood contour plots (see next section) are very
similar, which confirms the consistency of the two methods.

\section{Likelihood analysis}
\label{sec:likelihood}


By using the covariance matrices, we construct an \`a priori estimate on
cosmological parameter constraints from a cosmic shear survey. This
allows us to compare different survey geometries. As in Paper I, we use
the following figure-of-merit:
\begin{equation}
\chi^2(p) \equiv \sum_{ij} \left( \xi_i(p) - \xi^{\rm
t}_i \right) \Cov_{ij}^{-1} \left( \xi_j(p) - \xi^{\rm t}_j \right),
\label{chi-sqr}
\end{equation}
where the superscript $^{\rm t}$ denotes the fiducial model and $p$ is a
set of cosmological parameters which is tested against the fiducial
model, as specified in Sect.\ \ref{sec:cosm}. The summation indices
label the angular bins of the correlation functions. As noted in Paper
I, either $\xi_+$ or $\xi_-$ can be inserted for $\xi$ in
(\ref{chi-sqr}), in which case the corresponding covariance matrices
$\Cov_{++}$ or $\Cov_{--}$, respectively, have to be used for $\Cov$. The
resulting function is called $\chi^2_+$ or $\chi^2_-$ respectively. The
figure-of-merit $\chi^2_\tot$ combining all information is obtained
using the vector $\xi = \left( \xi_{+1}, \xi_{+2}, \ldots, \xi_{+N}, \,
\, \xi_{-1}, \ldots, \xi_{-N} \right)$ together with the block matrix
\begin{equation} {\rm Cov} = \left(\begin{array}{ll} {\rm Cov}_{++}   &
{\rm Cov}_{+-}\\[0.5cm] {\rm Cov}_{+-}^{\rm T} & {\rm
Cov}_{--}\end{array}\right) \label{cov-block}. \end{equation}

Analogously to (\ref{chi-sqr}), a figure-of-merit using the
$\average{M_{\rm ap}^2}$-statistics can be defined,
\begin{eqnarray} \lefteqn{\chi^2_{\rm E}(p) \equiv \sum_{ij} \left(
\average{M_{{\rm ap}}^2(\theta_i)}
- \average{M_{{\rm ap}}^2(\theta_i)}^{\rm t} \right)
  \Cov^{-1}_{ij}({\cal M}_+)} \nonumber \\ & & \times \left(
\average{M_{{\rm ap}}^2(\theta_j)} - \average{M_{{\rm
ap}}^2(\theta_j)}^{\rm t} \right), \label{chi-sqr-E} \end{eqnarray}
which is a likelihood measure using only the E-modes. Unfortunately,
$\chi_B^2$, using $M_\perp$ and thus testing the B-modes, cannot
properly be defined in this way. In the model used throughout in this
work, no B-modes are present, thus $\average{M_\perp^2} = 0$. The
correlation functions (\ref{xi-pm-def}) only depend on the E-mode
power spectrum; up to now, no models for the B-mode power spectrum have
been obtained. Predictions of the amplitude of intrinsic galaxy
alignment differ by orders of magnitudes and although some observational
results have been presented, stringent constraints have not yet been
obtained \citep[see][ and references therein]{brown02}.

\begin{figure*}[ht!]
\begin{center}
\includegraphics[width=12cm]{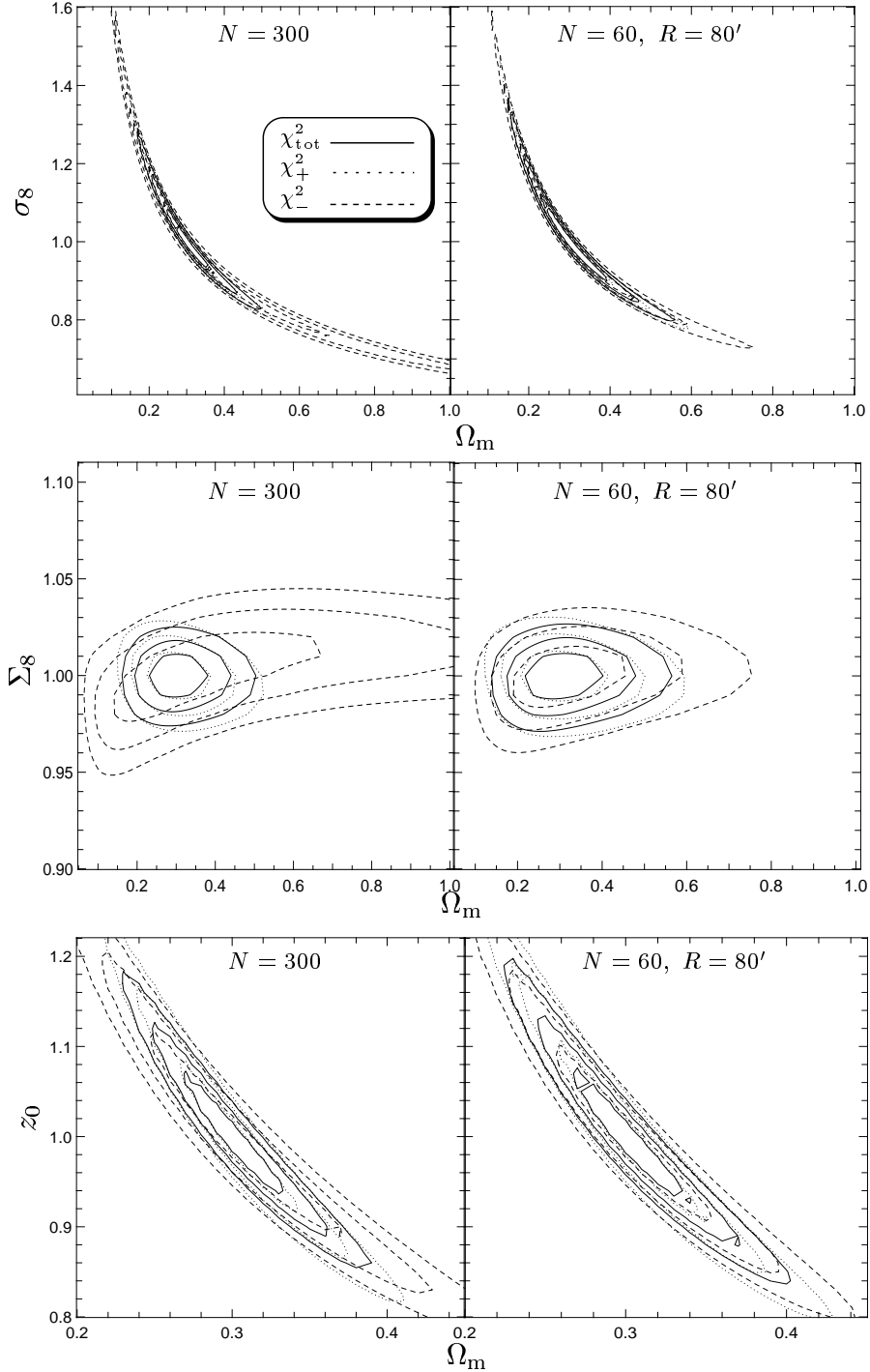}
\end{center}
\caption{1-$\sigma$, 2-$\sigma$ and 3-$\sigma$ confidence contours of
the figure-of-merit (\ref{chi-sqr}). Solid lines correspond to $\chi^2_{\rm tot}$,
dotted lines to $\chi^2_+$ and dashed lines to $\chi^2_-$. The left
panels show the results for the 300 uncorrelated images,
the right panels are for a $(N=60, R=80^\prime)$-patch geometry.
The parameter $\Sigma_8 \propto \sigma_8$ is defined in (\ref{Sigma_8}).}
\label{fig:contour1}
\end{figure*}

\begin{figure*}[ht!]
\begin{center}
\includegraphics[width=12cm]{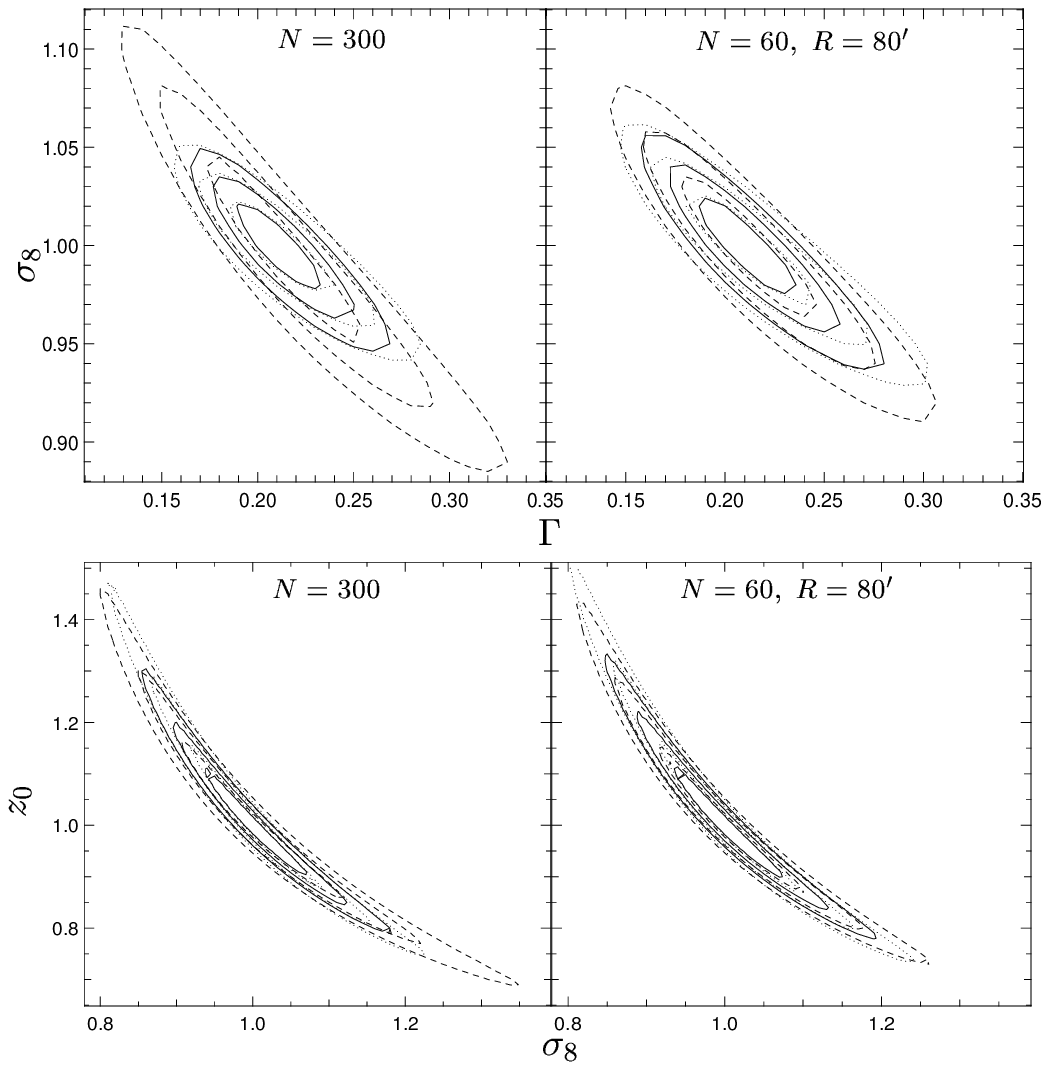}
\end{center}
\caption{Confidence levels of the figure-of-merit (\ref{chi-sqr}).
As in Fig.\ \ref{fig:contour1}, 1-$\sigma$, 2-$\sigma$ and 3-$\sigma$
contours are plotted; the solid lines display $\chi^2_{\rm tot}$, the
dotted and dashed lines correspond to $\chi^2_+$ and $\chi^2_-$, respectively.
The left panels show the results for the 300 uncorrelated images, the
right panels are for a $(N=60, R=80^\prime)$-patch geometry.}
\label{fig:contour2}
\end{figure*}

Two of the four parameters $\Omegam, \sigma_8, \Gamma$ and $z_0$ are
varied in a few combinations, while all others are kept fixed, with the
exception that a flat universe is assumed throughout.  In Figs.\
\ref{fig:contour1} and \ref{fig:contour2}, the expected likelihood
contours for the two extreme geometrical configurations regarding cosmic
variance are plotted: the uncorrelated images and the $(N=60,
R=80^\prime)$-patch geometry.

\subsection{$\Omegam$ and $\sigma_8$}

The strongest degeneracy between two parameters exists for $\Omegam$ and
$\sigma_8$, which is also expected from simulations and cosmic shear
measurements \citep[e.g.][]{vWM02}.
In order to compensate for the high elongation of the contours, we use
the combined parameter
\begin{equation}
\Sigma_8 \equiv \sigma_8 \left[ 0.41 + 0.59 \left(
\frac{\Omegam}{0.3} \right)^{-0.68} \right]^{-1}
\label{Sigma_8}
\end{equation}
which has been obtained in Paper I by fitting the minimum ``valley'' in the
($\Omegam$-$\sigma_8$)-plot.  Clearly, the $\chi^2_-$-contours are more
extended in the case of the uncorrelated images than for the patch
geometry. This is because $\xi_-$ contains much information on large
scales which is absent in the case of the uncorrelated images. In contrast,
the $\chi^2_+$-contours are tighter in this case than
for the patch geometry.

Furthermore, in both cases, the difference between $\chi^2_+$ and
$\chi^2_\tot$ are
small. Thus, most of the information concerning cosmology is contained in
$\xi_+$; the additional information coming from $\xi_-$ is relatively small.

\subsection{Other combinations}

There is also a strong degeneracy between other combinations of parameters, as
seen in Figs.\ \ref{fig:contour1} and \ref{fig:contour2}. In all cases, the
$\chi^2_{\rm tot}$- and $\chi^2_+$-contours are tighter for the uncorrelated
images, whereas the opposite is true for $\chi^2_-$. The patch survey geometry
yields constraints from $\xi_-$ compatible to those from $\xi_+$, in particular
when $z_0$ is one of the parameters; for the combination $\sigma_8$-$z_0$, the
$\chi_-$-contours are even tighter than the $\chi_+$-contours.




\subsection{Quadrupole moments}
\label{sec-quad}

For a more detailed analysis, a quantitative description of the
$\chi^2$-contour plots presented in the last section is needed. The quadrupole
moments of the underlying probability function can be used as a
measure of the surface of the contours. These are defined as
\begin{equation}
Q_{ij} \equiv \frac{\int \dd^2 p \, L(p_1, p_2) \left(p_i
- p_i^{\rm t}\right)
\left(p_j - p_j^{\rm t}\right)}{\int \dd^2 p \, L(p_1, p_2)}
\label{Qij}
\end{equation}
for $i,j = 1,2$.  The integration variables $p_1, p_2$ are the
parameters which are plotted on the axis in Figs.\ \ref{fig:contour1} and
\ref{fig:contour2}. $p_i^{\rm t}$ denotes the value of the parameter
of the fiducial set, $L \propto \exp(-\chi^2/2)$ is the likelihood
function.

The determinant of the quadrupole moment is then a measure of
the surface enclosed by the likelihood contours,
\begin{equation}
q \equiv \sqrt{\det Q_{ij}} = \sqrt{ Q_{11} Q_{22} - Q^2_{12}}.
\label{q}
\end{equation}
If the $\chi^2$-function was quadratic, the likelihood $L$ would be a
bivariate Gaussian. If in addition the two parameters $p_1$ and $p_2$ were
uncorrelated, $q \propto \Delta p_1 \cdot \Delta p_2$, where $\Delta
p_i$ is the variance of $p_i$.  In general, the smaller
the value of $q$, the tighter are the likelihood contours, and the
better the cosmological parameters can be constrained. For the
likelihood function $L$ in (\ref{Qij}), or equivalently, for $\chi^2$,
the three functions $\chi^2_+, \chi^2_-$ and $\chi^2_{\rm tot}$ can be
inserted. The resulting $q$'s will be refered to as $q_+, q_-$
and $q_{\rm tot}$ respectively. When (\ref{chi-sqr-E}) is used as the
figure-of-merit, the inferred contour surface measure is $\qE$.

\begin{figure}[ht!]
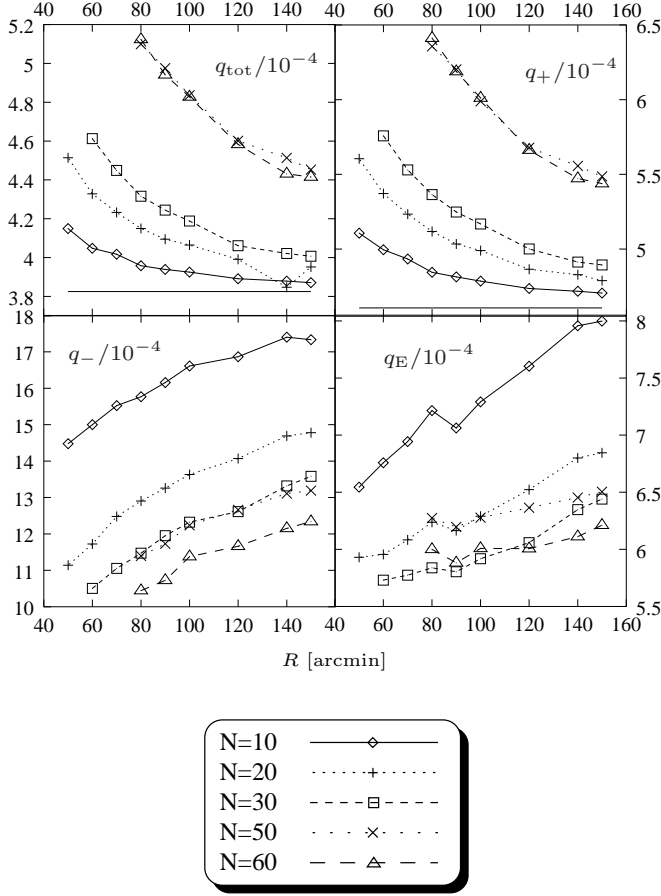

\begin{center}
\resizebox{\hsize}{!}{\input{OS.pstex_t}}
\medskip

\input{OSlegend.pstex_t}
\caption{The parameter $q$ (\ref{q}) for the
$\Omegam$-$\Sigma_8$-contours. On the $x$-axes, the patch radius $R$
in arc minutes is plotted, thus each point represents a patch geometry with
$N$ images in patches of radius $R$. The
horizontal line in the upper panels indicates the value for the 300 uncorrelated
images (omitted for $q_-$ and $\qE$, see also Table \ref{tab:q-single}).}
\label{fig:quad-OS}
\end{center}
\end{figure}

Values of $q$ for different geometries are given in Fig.\ \ref{fig:quad-OS}. The
likelihood contours of all three combinations of parameters considered here show
the same behaviour. Note that the values of $q$ only have a sensible meaning
when compared to each other for the same combination of cosmological parameters.

As expected, the tightest constraints
are obtained for $q_{\rm tot}$, as it combines the information of all
measurements.  $q_{\rm tot}$ is almost monotonically decreasing with
increasing patch radius $R$, and with decreasing number of images
per patch $N$.
This indicates that cosmic variance is the most crucial source of
errors: therefore, large patches and most notably a large number of directions
on the sky should be observed.

$q_+$ behaves similarly to $q_{\rm tot}$, but gives less tight
constraints on the parameters, as was already seen in the contour
plots.  The $q_-$-behaviour is opposite to $q_{\rm tot}$ and
$q_+$. The fact that $q_-$ increases towards small image numbers $N$
per patch is due to the weak dependence on cosmic variance for
$\Cov_{--}$. Further, small patches give better constraints,
indicating that dense sampling on medium scales is more important than
large-scale information.


The quadrupole moment measure $\qE$ of the $\average{M_{\rm ap}^2}$-statistics
shows a behaviour similar to $q_-$. This is because of the similarity between
$\xi_-$ and $\average{\Mapsq}$. However, while the $R$-dependence of $\qE$ is
quite monotonic, this is not the case for $N$. For most patch radii $R$, the
tightest constraints are archieved for $N=30$.

For the uncorrelated images, values of $q$ are given in Table
\ref{tab:q-single}; $q_{\rm tot}$ is smaller than for every patch
geometry.
As expected, $q_-$ is
larger than for any patch geometry, because $\xi_-$ is important on
medium scales which are not sampled by the uncorrelated images.


\begin{table}[tb]
\caption{Quantitative measures of the $\Omegam\,$-$\sigma_8$ likelihood contours for the
300 uncorrelated $13^\prime \times 13^\prime$ images. Table entries
are in units of $10^{-4}$.}
\begin{center}
\begin{tabular}{llll} \hline\hline
$q_{\rm tot}$ & $q_+$ & $q_-$ & $\qE$\\ \hline
3.825 & 4.606 & 26.53 & 12.31 \\
\hline
\end{tabular}
\end{center}
\label{tab:q-single}
\end{table}

In Sect. \ref{sec:fisher-more}, we take into account the simultaneous 
determination of three and four cosmological parameters.

\section{Fisher matrix analysis}
\label{sec:fisher}

The Fisher information matrix is defined as
\begin{equation}
F_{ij} = \average{\frac{\del^2 [-\ln L]}{\del p_i \del p_j}} =
\frac 1 2 \average{\frac{\del^2 \chi^2}{\del p_i \del p_j}}
\label{fisher-def}
\end{equation}
where $L$ is the likelihood function and $\chi^2$ is the
figure-of-merit (\ref{chi-sqr}) which depends on the parameters $p_i$
\citep[see e.g.][]{KS69, TTH97}.
The inverse of the Fisher matrix is a local measure of the curvature
at the minimum of $\chi^2$. Its eigenvalues and (pairwise orthogonal)
eigenvectors can be interpreted as the axes of an ellipsoid which
determines how fast the log-likelihood falls off the maximum in
different directions.  According to the Cram\'er-Rao inequality, the
smallest possible variance for any unbiased estimator of a parameter
$p_i$, if all parameters are to be estimated from the data, is
\begin{equation}
\sigma(p_i) = \sqrt{(F^{-1})_{ii}},
\label{sigma-fisher}
\end{equation}
thus defining a minimum variance bound.

Because we will later use the $\average{\Mapsq}$-statistics, we insert
(\ref{chi-sqr-E}) into (\ref{fisher-def}) and evaluate the equation at the
minimum of $\chi^2$ to get
\begin{equation}
F_{ij} = \sum_{kl} [\Cov^{-1}({\cal M}_+)]_{kl}\frac{\del \average{\Mapsq}_k}{\del p_i}
\frac{\del \average{\Mapsq}_l}{\del p_j}
\label{fisher2}
\end{equation}
Given the covariance matrices, we can easily calculate the minimum variance for the
cosmological parameters used in this analysis.

\subsection{Comparison with the likelihood}

\begin{figure}[ht!]
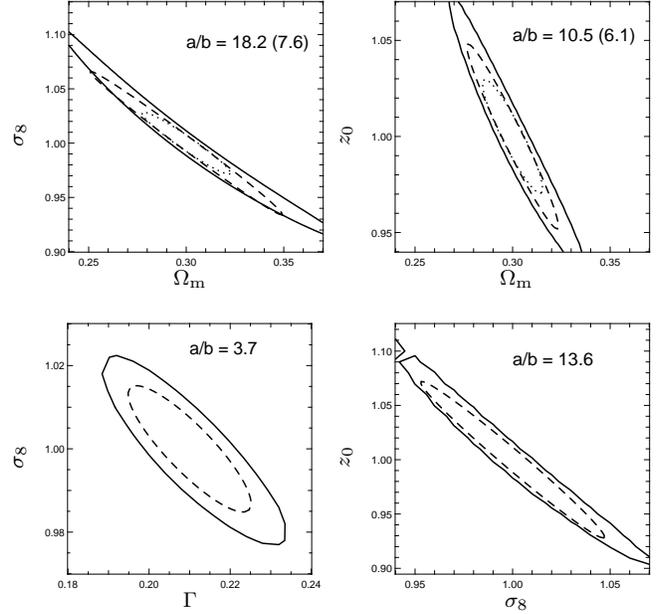

\input{13-1sigma-3.pstex_t}
\input{13-1sigma-2.pstex_t}

\bigskip

\input{13-1sigma-1.pstex_t}
\input{13-1sigma-5.pstex_t}

\caption{1-$\sigma$-likelihood contours (solid lines) using $\chi^2_{\rm tot}$
(\ref{chi-sqr}) compared with the 1-$\sigma$-error ellipse from the Fisher
matrix (\ref{fisher2}): the dashed ellipse is for a flat Universe (as it is the
case for the likelihood contours), the dotted one is for $\Omega_\Lambda=0.7$.
$a/b$ is the axis ratio of the ellipses (the case $\Omega_\Lambda=0.7$ is in
parentheses). The configuration is a survey of 300 uncorrelated images.}
\label{fig:combined} \end{figure}

First, we compare the minimum variance bound (\ref{sigma-fisher}) with
the likelihood contours from Sect.\
\ref{sec:likelihood}, using $\chi^2_{\rm tot}$. In this case, the Fisher information matrix
reduces to a $2\times2$ matrix. This comparison is shown in Fig.\
\ref{fig:combined}. As expected from the Cram\'er-Rao inequality,
the likelihood contours are larger than the 1-$\sigma$-ellipse from
the Fisher matrix. The orientation of the Fisher error ellipse
coincides with the likelihood shapes, i.e.\ the direction of the
minimal and maximal degeneracy of parameters is recovered. The
larger the degeneracy between two parameters, the larger is the
deviation between the local approximation by the Fisher matrix and the
likelihood function. For the case where the curvature is fixed (flat
Universe, $\Omegam + \Omega_\Lambda = 1$), the degeneracy is much
larger than for a fixed cosmological constant $\Omega_\Lambda = 0.7$.

\subsection{More parameters}
\label{sec:fisher-more}

Next, we calculate the minimum variance bound of three and more parameters out
of ($\Omegam, \sigma_8, \Gamma, \Omega_\Lambda, n_{\rm s}$) simultaneously,
corresponding to a full marginalization over these parameters. As seen in Fig.\
\ref{fig:quad-OS}, the aperture mass clearly gives less tight constraints than
the combined two-point correlation functions. However, this difference gets
smaller the more parameters are included, in some cases (for $\Gamma$ and
$n_{\rm s}$), the minimum variance bound is even smaller for $\average{\Mapsq}$.
Another advantage of the aperture mass is its ability to separate E- from
B-modes (see Sect.\ \ref{sec:rev}). Therefore, we will focus on this statistics
from now on. However, we must note that because of the local filtering of the
power spectrum, very large scales are not well sampled by $\average{\Mapsq}$.

\begin{figure}[ht!]
\resizebox{\hsize}{!}{\includegraphics{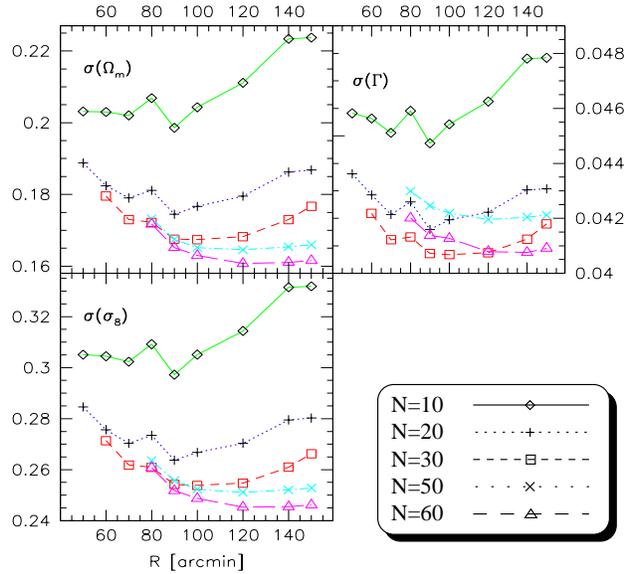}}
\caption{The minimum variance for the parameters $\Omegam, \Gamma$ and
$\sigma_8$ using (\ref{sigma-fisher}) and (\ref{fisher2}), for the
$\average{\Mapsq}$-statistics. $\Omegam, \Gamma$ and $\sigma_8$ are
assumed to be determined from the data, all other parameters are kept
fixed and a flat Universe is assumed. Each point in the plot
represents a patch geometry with $N$ images in patches of radius
$R$. The minimum variance bound for the uncorrelated images is given in Table \ref{tab:mvb}.}
\label{fig:fisherf-OGs}
\end{figure}

\begin{figure}[ht!]
\resizebox{\hsize}{!}{\includegraphics{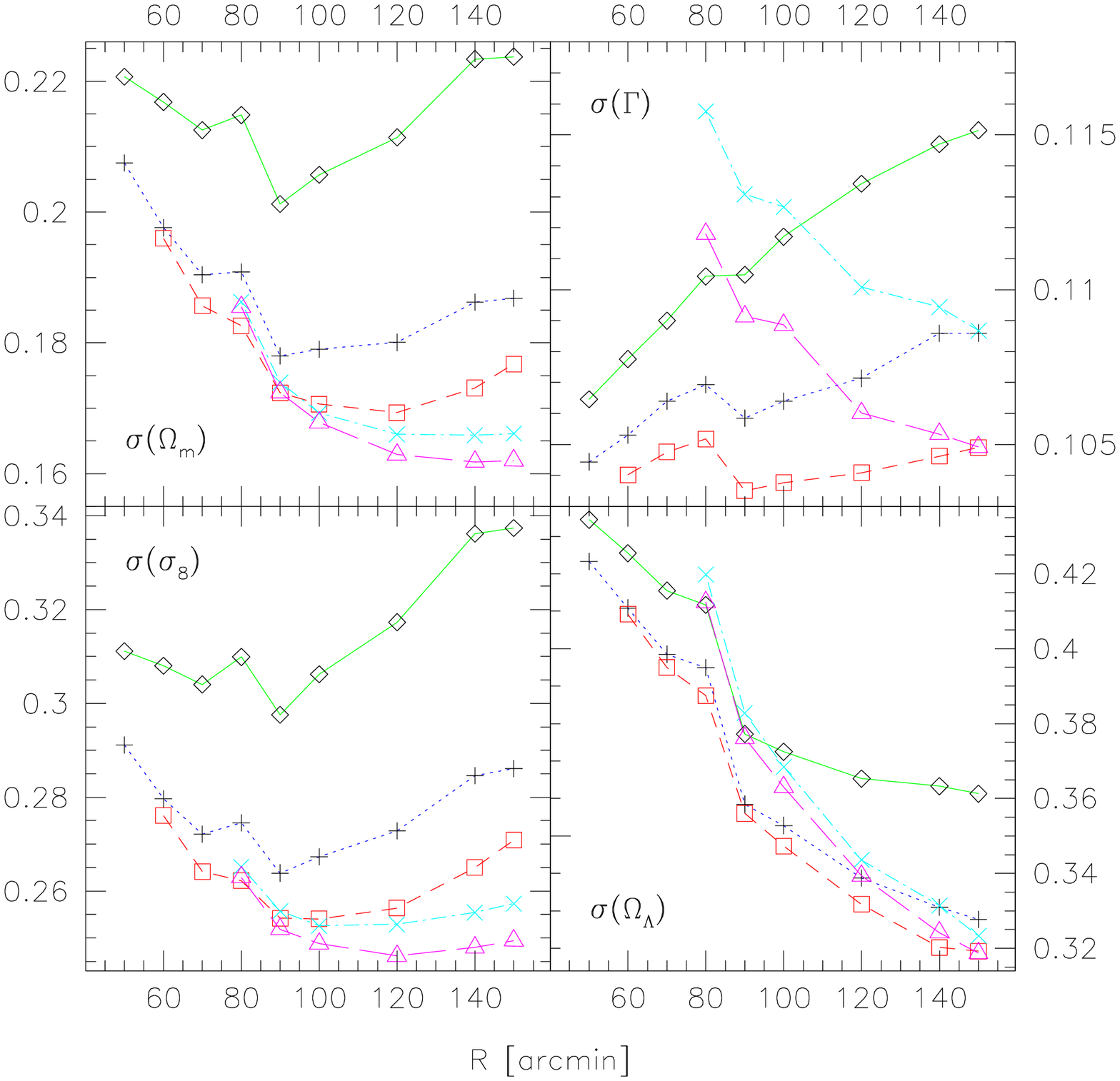}}
\caption{The minimum variance for the parameters $\Omegam, \Gamma, \sigma_8$ and
$\Omega_\Lambda$ using the $\average{\Mapsq}$-statistics. See Fig.\
\ref{fig:fisherf-OGs} for more details. The minimum variance bound for the uncorrelated images is given in Table \ref{tab:mvb}.}
\label{fig:fisher-OGsL}
\end{figure}

\begin{figure}[ht!]
\resizebox{\hsize}{!}{\includegraphics{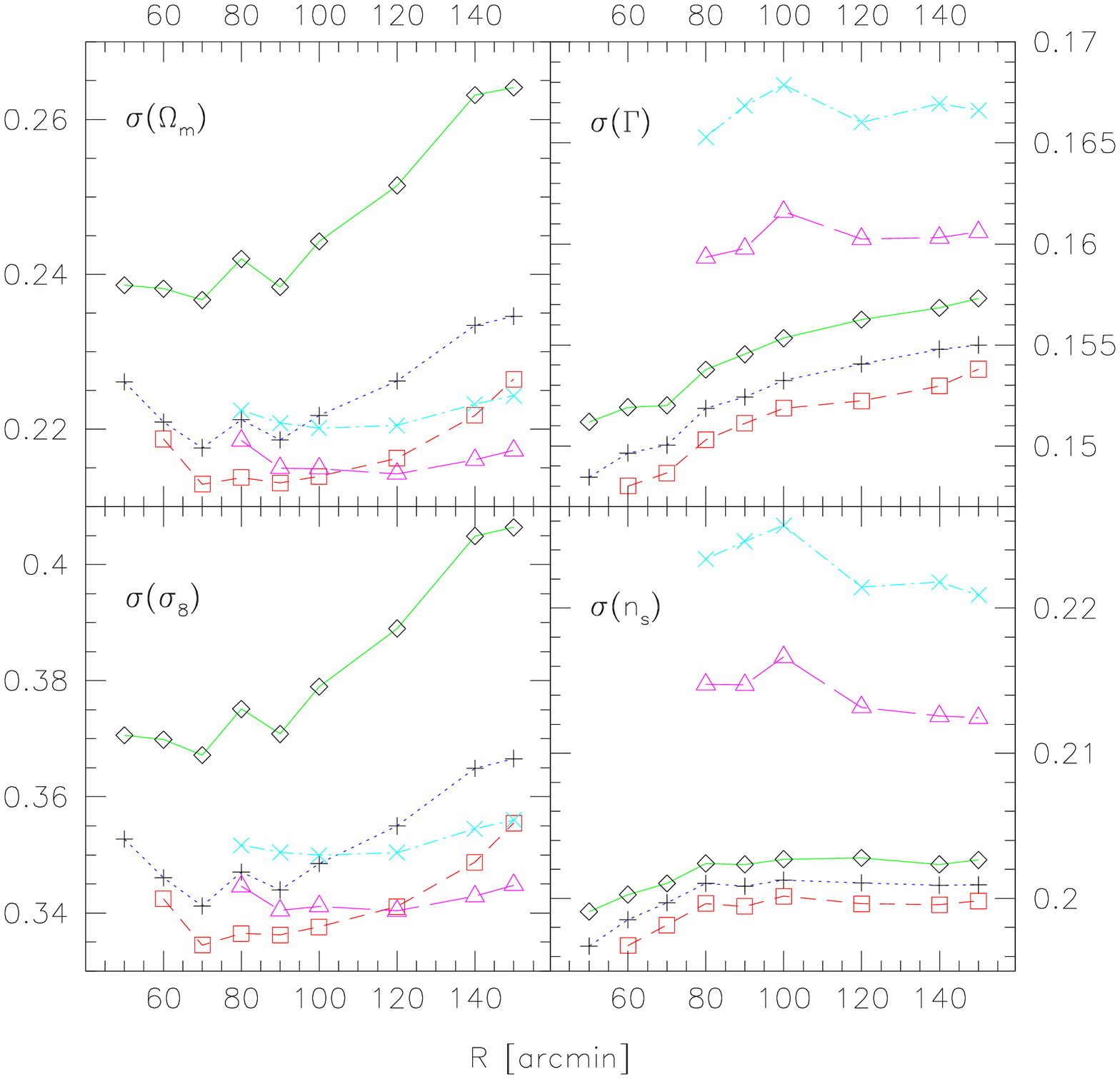}}
\caption{The minimum variance for the parameters $\Omegam, \Gamma, \sigma_8$ and
$n_{\rm s}$ using the $\average{\Mapsq}$-statistics. See Fig.\
\ref{fig:fisherf-OGs} for more details. The minimum variance bound for the uncorrelated images is given in Table \ref{tab:mvb}.}
\label{fig:fisherf-OGsn}
\end{figure}

Figs.\ \ref{fig:fisherf-OGs} - \ref{fig:fisherf-OGsn} show the minimum
variance for a different number of free cosmological parameters for the individual
patch geometries. The fixed parameters are set to the values given in
Sect.\ \ref{sec:cosm}. In the cases where $\Omega_\Lambda$ is not a free
parameter, the prior is a flat Universe ($\Omega_\Lambda = 1 - \Omegam$).

When taking into account three or more parameters, the uncorrelated
image configuration give very poor constraints on these parameters.
The minimum variance bound is in most cases more than double the value
of the least optimal patch geometry. The reason is that the lack of
large-scale information highly raises the degeneracy between
parameters. This cannot be compensated by the small cosmic variance.

\begin{table}
\caption{The minimum variance for several combinations of parameters, for the 300
uncorrelated image configuration. In each row, those parameters which
have an entry are assumed to be determined from the data, the other
parameters are fixed. The counterpart of the three rows for the patch
geometries are the Figs. \ref{fig:fisherf-OGs} -
\ref{fig:fisherf-OGsn}.}
\label{tab:mvb}
\begin{tabular}{lllll} \hline\hline
$\Omegam$ & $\sigma_8$ & $\Gamma$ & $\Omega_\Lambda$ & $n_{\rm s}$\\
\hline 0.53 & 0.76 & 0.09 & & \\ 0.53 & 0.76 & 0.14 & 0.64 & \\ 0.53 &
0.77 & 0.16 & & 0.20 \\ \hline
\end{tabular}
\end{table}

When adding $\Omega_\Lambda$ as a free parameter (compare
Fig. \ref{fig:fisherf-OGs} with Fig.\ \ref{fig:fisher-OGsL}), the
variance of the shape parameter increases by more than a factor of
two, whereas the variances of $\Omegam$ and $\sigma_8$ are only
slightly enhanced. 
For large patches, the variances of the three parameters are just rescaled,
whereas for small patches, the change is more complicated.
The constraint on the cosmological constant is very
poor, confirming a statement made by \cite{1997A&A...322....1B}.

When the spectral index $n_{\rm s}$ is added (Fig.\ \ref{fig:fisherf-OGsn}), the
minimum variance bound of $\Omegam$ and $\sigma_8$ again increase in
the same way as when adding $\Omega_\Lambda$, although by a greater amount.
The variance on $\Gamma$ changes completely, taking a similar
functional form on $R$ and $N$ as the variance on $n_{\rm s}$. The
reason for this is, that both parameters determine more or less the
shape of the power spectrum, whereas $\Omegam$ and $\sigma_8$
influence mainly its amplitude.

For each data point in Figs.\ \ref{fig:fisherf-OGs} -
\ref{fig:fisherf-OGsn}, corresponding to a survey with $N$ images in
$P = 300/N$ patches, only one realization of the random image
positions for each of
the $P$ patches was used. We produced some more realizations for two of
the patch geometries and found that the scattering of the minimum
variance bound is about one percent.

From Fig.\ \ref{fig:fisherf-OGs}, the best geometry is a survey with five
($N=60, R=120^\prime$)-patches. Considering Fig.\ \ref{fig:fisherf-OGsn}, a
configuration with $N=30$ and small $R$ yields the best minimum variance bounds.
For both cases, a survey with 30 images and a patch radius of around 100 arc
minutes seems to be a good choice. However, the patch radius only has a small
influence on the minimum variance bound, more important are the number of images
per patch and therefore the number of patches.

The difference in the minimum variance bound between individual patch
geometries can make up to 25 percent.



\section{Conclusions}
\label{sec:conc}

We calculated numerically the covariance matrices (\ref{Cov-terms} -
\ref{CovMap-def}) for the second-order estimators of cosmic shear
$\xi_\pm$ and $\average{\Mapsq}$, which were derived in Paper I, via a
Monte-Carlo-like technique. Galaxy positions were simulated for
various cosmic shear survey geometries of 14 square degree area. These
surveys consisted of a total of 300 images of size $13^\prime \times
13^\prime$ which were randomly distributed in patches on
the sky. A number of (semi-)random patch configurations were compared to a survey
consisting of 300 completely uncorrelated images. We performed several
analyses based on maximum likelihood and the Fisher information
matrix, enabling us to estimate the expected constraints on several
combinations of cosmological parameters. First, we assumed that only
two cosmological parameters are to be determined from the data, fixing
all the other parameters. In this case, using both two-point shear
correlation functions $\xi_+$ and $\xi_-$ in combination, the tightest
constraints were obtained for the uncorrelated image
configuration. Further, patch geometries with small cosmic variance
gave also good results. For the aperture mass statistics
$\average{\Mapsq}$, the best results came from a patch geometry with
$N=30$ images in 10 patches of radius $R=60^\prime$.  The uncorrelated
images could not compete with any patch geometry.

We then took into account three and four cosmological parameters out
of ($\Omegam, \sigma_8,
\Gamma, \Omega_\Lambda, n_s$).
The more parameters are assumed to be determined from the data, the
more important becomes large-scale information in order to 
resolve the near parameter degeneracies. Using the combined
$\xi_+$ and $\xi_-$, some of the patch geometries yield tighter
constraints than the uncorrelated image configuration. The aperture
mass is best applied to patches with $N=30$ images, the results are
nearly independent of the patch radius in most cases.

In most cases, the constraints obtained from the combined $\xi_+$ and
$\xi_-$ were tighter than those from $\average{\Mapsq}$. However, the
differences became smaller the more cosmological parameters were
included.

The differences between the individual patch geometries made up to 25
percent for the minimum variance bound on several parameters. Thus, a
25 percent improvement on the determination on cosmological parameters
can be obtained solely by choosing an appropriate geometry for a
future cosmic shear survey.

\begin{acknowledgements}
The authors want to thank Lindsay King, Ludovic van Waerbeke and Patrick
Simon for useful discussions, and the anonymous referee for helpful suggestions.
\end{acknowledgements}

\bibliographystyle{aa}
\bibliography{astro}

\begin{thebibliography}{36}
\expandafter\ifx\csname natexlab\endcsname\relax\def\natexlab#1{#1}\fi

\bibitem[{{Bacon} {et~al.}(2000){Bacon}, {R\'efr\'egier}, \&
  {Ellis}}]{2000MNRAS.318..625B}
{Bacon}, D.~J., {R\'efr\'egier}, A.~R., \& {Ellis}, R.~S. 2000, \mnras, 318,
  625

\bibitem[{Bardeen {et~al.}(1986)Bardeen, Bond, Kaiser, \& Szalay}]{bbks86}
Bardeen, J.~M., Bond, J.~R., Kaiser, N., \& Szalay, A.~S. 1986, \apj, 304, 15

\bibitem[{Bartelmann \& Schneider(2001)}]{BS01}
Bartelmann, M. \& Schneider, P. 2001, Phys.\ Rep., 340, 297

\bibitem[{Bennett {et~al.}(2003)Bennett, Halpern, Hinshaw,
  {et~al.}}]{Bennett03}
Bennett, C.~L., Halpern, M., Hinshaw, G., {et~al.} 2003, \apj, {Also}
  astro-ph/0302207

\bibitem[{{Bernardeau} {et~al.}(1997){Bernardeau}, {van Waerbeke}, \&
  {Mellier}}]{1997A&A...322....1B}
{Bernardeau}, F., {van Waerbeke}, L., \& {Mellier}, Y. 1997, \aap, 322, 1

\bibitem[{Brown {et~al.}(2002)Brown, Taylor, Hambly, \& Dye}]{brown02}
Brown, M.~L., Taylor, A.~N., Hambly, N.~C., \& Dye, S. 2002, \mnras, 333, 501

\bibitem[{Contaldi {et~al.}(2003)Contaldi, Hoekstra, \& Lewis}]{Contaldi03}
Contaldi, C.~R., Hoekstra, H., \& Lewis, A. 2003, \prl, submitted, {Also}
  astro-ph/0302435

\bibitem[{Gladders {et~al.}(2002)Gladders, Yee, McCarthy, Barrientos, Hoekstra,
  Hall, E., {et~al.}}]{G02}
Gladders, M.~D., Yee, H.~K.~C., McCarthy, P.~J., {et~al.} 2002, AAS, 201, 5906G

\bibitem[{Heymans \& Heavens(2003)}]{HH03}
Heymans, C. \& Heavens, A. 2003, \aap, 339, 711

\bibitem[{Hoekstra {et~al.}(2002)Hoekstra, Yee, Gladders, {et~al.}}]{hoe02}
Hoekstra, H., Yee, H., Gladders, M., {et~al.} 2002, \apj, 573, 55H

\bibitem[{Hu \& Tegmark(1999)}]{HuTeg99}
Hu, W. \& Tegmark, M. 1999, \apjl, 514, L65

\bibitem[{Jain {et~al.}(2000)Jain, Seljak, \& White}]{JSW00}
Jain, B., Seljak, U., \& White, S. 2000, \aap, 530, 547

\bibitem[{Jarvis {et~al.}(2003)Jarvis, Bernstein, Fischer, \& Smith}]{J03}
Jarvis, M., Bernstein, G.~M., Fischer, P., \& Smith, D. 2003, \aj, 125, 1014J

\bibitem[{{Kaiser}(1992)}]{1992ApJ...388..272K}
{Kaiser}, N. 1992, \apj, 388, 272

\bibitem[{{Kaiser}(1995)}]{1995ApJ...439L...1K}
---. 1995, \apjl, 439, L1

\bibitem[{{Kaiser}(1998)}]{1998ApJ...498...26K}
---. 1998, \apj, 498, 26

\bibitem[{Kaiser {et~al.}(2000)Kaiser, Wilson, \& Luppino}]{kaiser00}
Kaiser, N., Wilson, G., \& Luppino, G. 2000, astro-ph/0003338

\bibitem[{Kendall \& Stuart(1969)}]{KS69}
Kendall, M.~G. \& Stuart, A. 1969, The Advanced Theory of Statistics, Vol.~II
  (London: Griffin)

\bibitem[{King \& Schneider(2002)}]{KS02}
King, L. \& Schneider, P. 2002, \aap, 396, 411

\bibitem[{King \& Schneider(2003)}]{KS03}
---. 2003, \aap, 398, 23

\bibitem[{{Maoli} {et~al.}(2001){Maoli}, van Waerbeke, {Mellier}, {Schneider},
  {Jain}, {Bernardeau}, {Erben}, \& {Fort}}]{2001A&A...368..766M}
{Maoli}, R., van Waerbeke, L., {Mellier}, Y., {et~al.} 2001, \aap, 368, 766

\bibitem[{Peacock \& Dodds(1996)}]{pd96}
Peacock, J.~A. \& Dodds, S.~J. 1996, \mnras, 280, L19

\bibitem[{Press {et~al.}(1992)Press, Teukolsky, Flannery, \& Vetterling}]{nr}
Press, W.~H., Teukolsky, S.~A., Flannery, B.~P., \& Vetterling, W.~T. 1992,
  Numerical Recipes in C (Cambridge University Press)

\bibitem[{R\'efr\'egier {et~al.}(2002)R\'efr\'egier, Rhodes, \& Groth}]{R02}
R\'efr\'egier, A., Rhodes, J., \& Groth, E.~J. 2002, \apj, 572, L131

\bibitem[{Schneider(1996)}]{S96}
Schneider, P. 1996, \mnras, 283, 837

\bibitem[{{Schneider} {et~al.}(1998){Schneider}, van Waerbeke, {Jain}, \&
  {Kruse}}]{1998MNRAS.296..873S}
{Schneider}, P., van Waerbeke, L., {Jain}, B., \& {Kruse}, G. 1998, \mnras,
  296, 873

\bibitem[{Schneider {et~al.}(2002)Schneider, van Waerbeke, Kilbinger, \&
  Mellier}]{SvWKM02}
Schneider, P., van Waerbeke, L., Kilbinger, M., \& Mellier, Y. 2002, \aap, 396,
  1S, (Paper I)

\bibitem[{{Schneider} {et~al.}(2002){Schneider}, van Waerbeke, \&
  {Mellier}}]{2002A&A...389..729S}
{Schneider}, P., van Waerbeke, L., \& {Mellier}, Y. 2002, \aap, 389, 729

\bibitem[{{Scoccimarro} {et~al.}(1999){Scoccimarro}, {Zaldarriaga}, \&
  {Hui}}]{1999ApJ...527....1S}
{Scoccimarro}, R., {Zaldarriaga}, M., \& {Hui}, L. 1999, \apj, 527, 1

\bibitem[{Smail {et~al.}(1995)Smail, Hogg, Yan, \& Cohen}]{Smail95}
Smail, I., Hogg, D.~W., Yan, L., \& Cohen, J.~G. 1995, \apjl, 449, L105

\bibitem[{Spergel {et~al.}(2003)Spergel, Verde, Peiris, Komatsu, R., Bennett,
  {et~al.}}]{Spergel03}
Spergel, D.~N., Verde, L., Peiris, H.~V., {et~al.} 2003, \apj, in press, {Also}
  astro-ph/0302209

\bibitem[{Tegmark {et~al.}(1997)Tegmark, Taylor, \& Heavens}]{TTH97}
Tegmark, M., Taylor, A., \& Heavens, A. 1997, \apj, 480, 22

\bibitem[{van Waerbeke {et~al.}(2000)van Waerbeke, {Mellier}, {Erben},
  {Cuillandre}, {Bernardeau}, {Maoli}, {Bertin}, {McCracken}, {Le F{\` e}vre},
  {Fort}, {Dantel-Fort}, {Jain}, \& {Schneider}}]{2000A&A...358...30V}
van Waerbeke, L., {Mellier}, Y., {Erben}, T., {et~al.} 2000, \aap, 358, 30

\bibitem[{van Waerbeke {et~al.}(2002)van Waerbeke, Mellier, Pell\'o, Pen,
  McCracken, \& Jain}]{vWM02}
van Waerbeke, L., Mellier, Y., Pell\'o, R., {et~al.} 2002, \aap, 393, 369V

\bibitem[{van Waerbeke {et~al.}(2001)van Waerbeke, {Mellier}, Radovich, Bertin,
  Dantel-Fort, McCracken, Le Fèvre, Foucaud, Cuillandre, Erben, Jain,
  Schneider, Bernardeau, \& Fort}]{vWMR01}
van Waerbeke, L., {Mellier}, Y., Radovich,  ., {et~al.} 2001, \aap, 374, 757V

\bibitem[{{Wittman} {et~al.}(2000){Wittman}, {Tyson}, {Kirkman},
  {Dell'Antonio}, \& {Bernstein}}]{2000Natur.405..143W}
{Wittman}, D.~M., {Tyson}, J.~A., {Kirkman}, D., {Dell'Antonio}, I., \&
  {Bernstein}, G. 2000, Nature, 405, 143

\end{thebibliography}

\end{document}